\input phyzzx.tex
\tolerance=1000
\voffset=-0.0cm
\hoffset=0.7cm
\sequentialequations
\def\rl{\rightline}

\def\t1{{\tilde 1}}

\def\f{\phi}
\def\t{\theta}
\def\p{\psi}

\REF{\DTE}{E. Halyo, Phys. Lett. {\bf B387}  (1996) 43, hep-ph/9606423; P. Binetruy and G. Dvali, Phys. Lett. {\bf B388}  (1996) 241, hep-ph/9606342;
E. Halyo, Phys. Lett. {\bf B454} (1999) 223; hep-ph/9901302.}
\REF{\BRA}{G. Dvali and S. H. Tye, Phys. Lett. {\bf B450} (1999), hep-ph/9812483.}
\REF{\KUT}{D. Kutasov and A. Giveon, Rev. Mod. Phys. {\bf 71} (1999) 983, hep-th/9802067.}
\REF{\CHA}{A. D. Linde, Phys. Lett. {\bf B129} (1983) 177; Phys. Lett. {\bf B175} (1986) 395.}
\REF{\HYB}{A. D. Linde, Phys. Lett. {\bf B259} (1991) 38; Phys. Rev. {\bf D49} (1994) 748.}
\REF{\DVA}{S. Alexander, hep-th/0105032; G. Dvali, Q. Shafi and S. Solganik, hep-th/0105203; C. P. Burgess et al., hep-th/0105204.}
\REF{\SEN}{A. Sen, Int. J. Mod. Phys. {\bf A14} (1999) 4061, hep-th/9902105 and references therein.}
\REF{\BRO}{J. H. Brodie, hep-th/0101115 and references therein.}
\REF{\QUI}{E. Halyo, hep-ph/0105216.}
\REF{\EH}{E. Halyo, Phys. Lett. {\bf B461} (1999) 109, hep-ph/9905244; JHEP 9909 (1999) 012, hep-ph/9907223.}

\singlespace
\rl{SU-ITP-01-}
\rl{hep-ph/0105341}
\rl{\today}
\pagenumber=0
\normalspace
\medskip
\bigskip
\titlestyle{\bf{ Inflation from Rotation}}
\smallskip
\author{ Edi Halyo{\footnote*{e--mail address: vhalyo@.stanford.edu}}}
\smallskip
\centerline {Department of Physics}
\centerline{Stanford University}
\centerline {Stanford, CA 94305}
\centerline{and}
 \centerline{California Institute for Physics and Astrophysics}
\centerline{366 Cambridge St.}
\centerline{Palo Alto, CA 94306}
\smallskip
\vskip 2 cm
\titlestyle{\bf ABSTRACT}

We obtain brane inflation in configurations with Dp and D(p-2) branes which are far apart and almost perpendicular to each other. Chaotic brane
inflation occurs when the
branes can only rotate with the inflaton given by the angle between the branes. When the branes can also move hybrid brane inflation occurs where the inflaton is the distance
between the branes and the trigger field is related to the angle. In both models we find a spectral index very close to one. Both types of inflation require $M_s \sim 10^{16-17}~GeV$
with one large radius of size $1/R \sim 10^{12-15}~GeV$.

\singlespace
\vskip 0.5cm
\endpage
\normalspace

\centerline{\bf 1. Introduction}
\medskip

Inflation occurs at very high energies and therefore must be embedded in a theory such as supergavity (or even possibly string
theory). This was hard to accomplish for a long time due to the large generic inflaton mass in supergravity models with F--term supersymmetry breaking.
This problem was finally circumvented by using D--term supersymmetry breaking as in D--term inflation[\DTE].

In this paper, we consider another possibility for inflation in supergravity which uses branes[\BRA].
Specifically, we consider a D(p-2) brane almost perpendicular to a Dp brane at a large distance (compared to the string scale). It is well--known that this configuration of branes
is supersymmetric when the branes are perpendicular[\KUT]. For a small deviation from this configuration supersymmetry is broken; a potential on the brane is generated and inflation can take
place. Inflation ends when
the brane configuration reaches the supersymmetric one. We show that depending on reasonable assumptions about the brane setup one can obtain either chaotic[\CHA] or hybrid
[\HYB]inflation
on D-branes which satisfy all observational constraints. For chaotic brane inflation, the branes are fixed (e.g. on orbifold planes) in space but can rotate. The inflaton is the
the scalar field on the brane which parametrizes the angle. Hybrid brane inflation occurs when at least one of the branes can move. In this case, the inflaton parametrizes
the distance between the branes whereas the trigger field is related to the relative angle.
Both scenarios can be effectively discussed in supergravity since for large distances supergravity is a very good approximation to string theory.

Very recently, a number of other works appeared which consider inflation in a brane--antibrane setting which are similar to our work in spirit[\DVA].
In these models, the branes and antibranes eventually fall
on each other and annihilate. This process is not well--understood in string theory; for example depending on the tachyon potential brane--antibrane annihilation may lead to lower
dimensional nonsupersymmetric branes or to a closed string vacuum[\SEN]. On the other hand, in our case, the branes eventually become perpendicular and the configuration becomes supersymmetric.
As a result, we can precisely follow the evolution of the branes until the very end. The main difference between our models and those with brane--antibrane configurations is in the nature
of the potential generated on the
branes. Models in ref. [\DVA] result in an attractive potential between the branes with an additional constant term given by the brane tension. In our models, due to the special
configuration the brane potential is repulsive which leads to the different inflationary scenarios.
Moreover, we eliminate the brane tension (cosmological constant) by placing the brane on an orientifold plane with negative tension.

This letter is organized as follows. In section 2 we obtain the potential on the brane from interbrane interactions in supergravity. In section 3, we use this to obtain
chaotic inflation on the brane. In section 4 we discuss the hybrid inflation on the brane. Section 5 contains our conclusions and a discussion of our results.

\bigskip
\centerline{\bf 2. D-Branes at an Angle}
\medskip

The prototypical brane configuration we will consider is given by a $Dp$ brane on top of an orientifold plane with another $D^p{\prime}$ brane at an angle
and at a large distance. The orientifold which has negative tension is needed to cancel the positive brane tension (i.e. to give vanishing cosmological constant on the brane)
and the second brane
is at an angle in order to break supersymmetry and generate the inflaton potential. Here we assume that the space transverse to the $Dp$ brane $T^{9-p}$ is orbifolded
so that on the brane there is only $N=1$ supersymmetry before it is broken by the relative angle.
Consider a $Dp$ and a $D^p{\prime}$ brane almost perpendicular and separated by a distance $r$. (We ignore the orientifold plane, $Op$ in the following since
all it does is to cancel the constant term in the potential.) We will compute the potential on the $Dp-Op$ pair created by the presence
of the $Dp^{\prime}$ brane by using supergravity[\BRO]. This is justified because supergravity is a very good approximation to the full string
theory when the branes are very far from each other (compared to the string scale).

The metric generated by the $Dp^{\prime}$ brane is given by
$$ds^2=h(r)^{-1/2}dx^2_{par}+h(r)^{1/2}dx^2_{perp} \eqno(1)$$
where the subscripts $par$ and $perp$ denote the directions parallel and perpendicular to the brane world--volume and
$$h(r)=1+g_s({1 \over {M_s r}})^{7-p^{\prime}} \eqno(2)$$
with $M_s$ the string scale and $g_s$ the asymptotic value of the string coupling.
The dilaton profile generated by this brane is
$$e^{-2D}=h(r)^{(p^{\prime}-3)/2} \eqno(3)$$

The above supergravity background gives us the potential on the $Dp$ brane when substituted into the Born--Infeld action
$$S_p={M_s^{p+1} \over g_s}   \int d^{p+1}y e^{-D} \sqrt{det g_{\mu \nu}}   \eqno(4)$$
where $g_{\mu \nu}$ is the pullback of the metric in eq. (1) to the $p+1$ dimensional world--volume.
Using eqs. (1-4) we get the potential
$$V_p(r)={M_s^{p+1} \over g_s} h(r)^{(p^{\prime}-3)/4} h(r)^{-(p+1)/4}(1+O((\partial X)^2)+ \ldots  \eqno(5)$$
In the expansion of the potential the first term is the $Dp$ brane tension which is cancelled by the negative and equal orientifold tension.
The next terms in the expansion are
$$V_p(r)=M_s^{p+1} \left ({{p^{\prime}-p-3}\over 4} \right) ({1 \over {M_s r}})^{7-p^{\prime}} +\ldots)  \eqno(6)$$

Now we consider the case with $p=p^{\prime}-2$ in which one of the branes is rotated relative to the other by an angle $\pi/2+\t$, i.e. the branes are almost perpendicular to
each other.
In this case there is a repulsive potential between the branes. The full potential in eq. (5) becomes[\BRO]
$$V_p(r)= {M_s^{p+1} \over g_s} h(r)^{(p^{\prime}-p-3)/4} (sin^2 \t h(r)^{1/2}+cos^2 \t h(r)^{-1/2})^{1/2} \eqno(7)$$
The above potential vanishes when the relative angle between the branes becomes $\pi/2$ since this configuration is supersymmetric.
Including the negative orientifold tension, for small $\t$ and large distances $r>>M_s^{-1}$ the potential in eq. (7) becomes
$$V_p(r) \sim M_s^{p+1}(\t^2({1 \over {M_s r}})^{7-p^{\prime}}+ \ldots) \eqno(8)$$
The distance between the branes is parametrized by the scalar field $\f= r M_s^2 $ whereas the angle is parametrized by $\p= 2\t r M_s^2$.

\bigskip
\centerline{\bf 3. Chaotic D-Brane Inflation from Rotation}
\medskip

In this section we assume that the two types of branes are fixed in space (e.g. by orbifolds and/or or orientifolds) so that they cannot move but can rotate. In this case, $\f$ has a value fixed
by the distance $r$  between the branes and is not dynamical. The field that parametrizes the angle, $\t$ is the inflaton with a potential
$$V_p(\t) \sim M_s^{p+1}(\t^2({1 \over {M_s r}})^{7-p^{\prime}}+ \ldots) \eqno(9)$$
Consider, now the case with $p=3$ and $p^{\prime}=5$ which are almost perpendicular, i.e. at an angle $\pi/2+\t$. (We assume that we live on the D3 brane.)
The potential becomes
$$V(\t) \sim {M_s^2 \over r^2} \t^2 \eqno(10)$$
The dimensionless angle $\t$ is related to a dimension one field $\p$ by $\p= 2 \t M_s^2 r$. The potential becomes
$$V(\p) \sim  \p^2 {1 \over M_s^2 r^4} \eqno(11)$$
where $r$ is fixed and large.
This is the potential for chaotic inflation[\CHA]. The inflaton $\p$ satisfies the equation of motion
$${\ddot {\p}}+3H {\dot \p}+V^{\prime}=0 \eqno(12)$$
For inflation to occur the potential has to satisfy the slow--roll condition $H^2>>m_{\p}^2$ where the Hubble constant is
given by $H=V/M_P^2$. This condition translates into $\p \sim \t M_s^2 r>M_P$. Assuming only one compactified dimension  with radius $R$ larger than $M_s^{-1}$ (and the others
of order $M_s^{-1}$) we have
$R M_s^3 \sim M_P^2$. Therefore, the slow--roll condition becomes $\t r>R^{1/2} M_s^{-1/2}$. We see that, even for small angles
inflation occurs if the distance between the branes is large enough, e.g. for $r \sim R > 10^3 M_s^{-1}$ (which means $M_s \sim 0.1 M_P$).

The other condition for inflation is given by
$$\epsilon=M_P^2 \left(V^{\prime} \over V \right)<<1 \eqno(13)$$
This is automatically satisfied for $\p >>M_P$ or $r \sim 10^3 M_s^{-1}$. Inflation ends when one or both these conditions fail to hold, i.e.
when $\t= \t_c \sim (rM_s)^{-1/2} \sim 0.03$.
The number of e--folds during inflation is given by
$$N \sim \int_{\p_c}^{\p_i} M_P^{-2} (V/V^{\prime}) \eqno(14)$$
We find that $N \sim (\p_i^2-\p_c^2)/M_P^2$. In terms of the angle, $N \sim M_s^4 r^2 (\t_i^2-\t_c^2)/M_P^2 \sim r M_s(\t_i^2-\t_c^2)$.
For $r \sim 10^3 M_s^{-1}$ required by the slow--roll conditions, we find that we can easily get more than 60 e--foldings for small angles, e.g. $\t_i \sim 0.3$. Note that
for larger angles $N$ can become much larger, of order $10^3$.

The magnitude of the density perturbations from COBE data require
$${\delta \rho \over \rho} \sim \left(V^{3/2} \over {M_P^3 V} \right) \sim 2 \times 10^{-5} \eqno(15)$$
This gives a constraint on $\t_i$, $\t_i^2/M_Pr \sim 10^{-5}$. Using $r \sim 10^4 M_P^{-1}$ we find $\t_i \sim 0.3$ a value consistent with $N \sim 100$. Note that a change
by a factor of 5 in the ratio $M_s/M_P$ would completely lift the constraint on $\t$, i.e. the bound becomes $\t \leq 1$. This requires a larger compactification radius
$R \sim 10^5 M_s^{-1}$ but gives a larger number ($\sim 10^3$) of e--foldings.

The spectral index is given by
$$n \sim 1-{2 \over N} \eqno(16)$$
Since we can make $N$ as large as $10^3$, chaotic brane inflation can accomodate an index very close to 1.

Inflation ends when $\t$ reaches its critical value $\t=\t_c \sim 0.03$ and starts to roll down fast. The damped oscillations of $\p$ around its minimum reheats the universe.
Since $\p$ is a specific combination of world--volume degrees of freedom like all other brane fields, we will assume that its interactions with (fermionic) matter on the brane is of
Yukawa type with a coupling $g \sim 1$. In that case the decay rate is
$$\Gamma_{\t} \sim {g^2 m_{\t} \over 8 \pi} \eqno(17)$$
which using
$$T_R \sim 10^{-1} \sqrt{\Gamma_{\t} M_P} \eqno(18)$$
gives $T_R \sim 10^{14}~GeV$.

From the bulk point of view inflation occurs when the branes are far apart and at an angle slightly different than $\pi/2$.
During inflation one of the branes is rotating (the angle is decreasing) very slowly
decreasing the vacuum energy with it. After some time $H \sim m_{\t}$ and the brane starts to rotate faster and inflation ends. The brane rotates back to being parallel with
the other one and starts to perform damped oscillations around the parallel configuration. This is the era of reheating. Finally it stops due to the friction term in eq. (12).

We find that chaotic brane inflation can be easily obtained from a Dp--D(p-2) brane configuration in which the branes are almost perpendicular and at a large distance. The
inflaton is given by the field that parametrizes the relative angle between the branes.
In order to accomplish this we
require $M_s/M_P \sim 0.1$ and one large dimension with $1/R \sim 10^{15}~GeV$. Above, for simplicity we considered the case with $p=3$ and $p^{\prime}=5$ where we live on the
$D3$ brane. Our scenario applies equally well to other configurations with $p=p^{\prime}-2$; however, in those cases the parameters of the model are slightly different
in order to accomodate the observed values of $n$ and $\delta \rho/\rho$.

\bigskip
\centerline{\bf 4. Hybrid D-Brane Inflation from Rotation}
\medskip

We will now consider the same brane configuration in the previous section but
relax the assumption that both branes are fixed so that at least one of them can move.
The potential now contains two fields, $\f$ and $\p$ and is given by
$$V_p(\p,\f) \sim M_s^{p-p^{\prime}+8}(\p^2({1 \over {\f}})^{7-p^{\prime}}) \eqno(19)$$
This results in hybrid inflation in which the inflaton is $\p$ (which parametrizes the distance between the branes) and the trigger field is $\f$ (which parametrizes
the relative angle between them). Exactly the same potential has been used recently to describe a model of hybrid quintessence which comes to an end and therefore does not
suffer from problems related to cosmological horizons[\QUI].
We consider as before the case with a D5 brane almost perpendicular to a D3 brane (with a relative angle $\pi/2+\t$) so that the inflaton potential becomes
$$V(\p,\f) \sim M_s^6 {\p^2 \over \f^4} \eqno(20)$$
Before inflation both fields have large values, $\f \sim \p > M_P$. We have for their masses
$$m_{\p}^2 \sim  {M_s^6\over {\f}^4} \eqno(21)$$
and
$$m_{\f}^2 \sim  M_s^6{\p^2 \over {\f}^6} \eqno(22)$$
Therefore the slow--roll conditions $H^2>>m_{\p}^2$, $H^2>>m_{\f}^2$ and also $\epsilon<<1$ are satisfied for $\t r_i >10 M_P/M_s^2$ as in the previous section. Again, if we take
$M_s \sim 0.1 M_P$ and for one large dimension we find that the fields are slowly rolling if $r_i \sim R >10^3 M_s^{-1}$.
During this era of infaltion, $\f$ and $\p$
are slowly rolling down their potential ($\f$ to larger values and $\p$ to smaller values). As $\p$ rolls
to smaller values at some critical value $\p_c \sim M_P$ or $(r \t)_c \sim 100 M_s^{-1}$ we obtain
$H \sim m_{\p}$. Then $\p$ starts to roll down fast and inflation ends. The number of e--folds is given by eq. (14)
which assuming that $\p$ is constant during inflation gives
$N \sim (\f_i^2-\f_c^2)/M_P^2 \sim M_s^4 M_P^{-2}(r_i^2-r_c^2)$. Once again for $r_i > 10 M_P/M_s^2$ this easily gives $N \sim 100$ or more.
The magnitude of density perturbations is given by (up to an irrelevant sign)
$${\delta \rho \over \rho} \sim \left(V^{3/2} \over {M_P^3 V} \right) \sim  \t \left(M_s \over M_P \right)^3  \sim 2 \times 10^{-5} \eqno(23)$$
In order to keep the value of $r \t$ above the critical value and satisfy the above constraint, we need a smaller ratio $(M_s/M_P) \sim 0.02$ which requires
$R \sim 10^5 M_s^{-1}$. Then, the correct magnitude of the density perturbatons is obtained for $\t \sim 0.1$. This forces a very large number of e--foldings, e.g. $N \sim 10^4$.
The spectral index again is given by eq. (16)
and in this case has to be very close to 1 due to the large $N$.
After inflation ends due to the fast rolling of $\p$, $\p$ reaches its minimum at vanishing VEV and starts to oscillate about it. The damped oscillations of the trigger field $\t$
reheat the universe up to $T_R \sim 10^{12}~GeV$.

Seen from the ten dimensional bulk point of view, the evolution during and after inflation is as follows. At the beginning of inflation, the two branes are far from
each other and almost perpendicular. During inflation the distance increases and angle decreases very slowly. As the trigger field, $\p$ starts
to roll down its potential one of the branes starts to rotate fast and inflation ends. Eventually the brane starts to perform damped oscillations (reheating era)
around the parallel configuration and stops. However, $\f$ is still slowly increasing due to
its kinetic energy which means that the branes are slowly separating. This kinetic energy will decrease very fast with time compared to matter
energy density and will become negligible. From the bulk point of view, using
eq. (12) we see that the brane will come to a stop due to the friction which arises from the Hubble term.

We see that when the branes can move the brane configuration we considered in this paper leads to hybrid inflation. The inflaton is given by the distance between the branes
whereas the trigger field is given by the relative angle. In this case we find that we need a small string scale with $M_s/M_P \sim 0.02$ and one rather large compactification
radius $1/R \sim 10^{12}~GeV$. In this case, the spectral index has to be extremely close to 1.
Our results for hybrid brane inflation are also valid for branes with different dimensions as long as $p=p^{\prime}-2$ with a small change of
parameters such as $M_s$ and $R$.

\bigskip
\centerline{\bf 5. Conclusions and Discussion}
\medskip

In this paper, we showed that brane configurations with two codimension two branes almost perpendicular to each other and at a large distance can lead to either chaotic or
hybrid inflation. We found that for chaotic brane inflation to accomodate cosmological data the string scale must be about a factor of 10 smaller than Planck scale with one large
dimension, i.e. $1/R \sim 10^{15}~GeV$. The spectral index satisfies $0.99<n<1$. For brane hybrid inflation data requires $M_s \sim 0.02 M_P$ and one rather large
dimension with $1/R \sim 10^{12}~GeV$. In this case, $n$ has to be extremely close to 1.
A robust prediction of the models we discussed above is the requirement for only one large dimension which accounts for the necessary small hierarchy between $M_s$ and $M_P$.
It is easy to see from eq. (14) that for more than one large dimension one cannot obtain enough e--foldings during inflation in either model.

The difference between our models and those with branes--antibranes which appeared recently is in the scalar potential on the branes which ar obtained from the interbrane
interactions in supergravity. our brane configurations are such that the interactions are repulsive leading to a positive potential for the inflaton (without the constant term
due to the brane tension which is cancelled by an orientifold). In models of ref. [\DVA], the interactions are attractive and the constant term is required to get an overall
positive potential. One can easily imagine brane configurations with elements from both scenarios. For example, there may be inflation in brane--antibrane configurations which are
at an angle or in in brane--brane configurations (at an angle) with an attractive interbrane potential. These may lead to inflationary models with more natural parameters than the
ones in this paper.

A curious fact about our models is their relation with D--term inflation. If (or when) the branes are closer to each other than the string size supergravity breaks down and the
physics is described by gauge theories living on the branes. In this case, the relative angle between the branes is described by an anomolous D--term[\KUT] just as in D--term inflation.
Thus, it seems that the above inflationary scenarios with branes at an angle are continously connected to D--term inflation on branes [\EH]with the interbrane distance the interpolating
parameter.

\bigskip
\centerline{\bf Acknowledgements}

I would like to thank John Brodie for a very useful discussion.

\vfill

\refout

\end
\bye